\documentclass[10pt,letterpaper]{article}
\usepackage{opex3}
\usepackage{mathptmx}
\usepackage{color}
\usepackage{graphicx}
\usepackage{color}
\usepackage{cite}

\begin{document}

\title{Focused electromagnetic doughnut pulses and their interaction with interfaces and nanostructures}

\author{Tim Raybould$^{1,*}$, Vassili Fedotov$^{1}$, Nikitas Papasimakis$^{1}$, Ian Youngs$^{2}$ and Nikolay Zheludev$^{1,3}$}

\address{$^1$Optoelectronics Research Centre \& Centre for Photonic Metamaterials, University of Southampton, UK\\
$^2$DSTL, Salisbury, UK\\$^3$Centre for Disruptive Photonic Technologies, Nanyang Technological University, Singapore }

\email{$^*$T.A.Raybould@soton.ac.uk} 



\begin{abstract}
“We study the propagation properties and light-matter interactions of the “\textit{focused doughnut}” pulses, broadband, single-cycle electromagnetic perturbations of toroidal topology first described by Hellwarth and Nouchi in 1996. We show how “focused doughnuts” are reflected  and refracted at planar metallic and vacuum-dielectric interfaces leading to complex distortions of the field structure. We also identify the conditions under which these toroidal pulses excite dominant dynamic toroidal dipoles in spherical dielectric particles.”
\end{abstract}




\section{Introduction}

The homogenous Maxwell's equations describe the behavior of electromagnetic radiation in free space. Infinite energy plane waves of the form $Ae^{i(kr-\omega t)}$ are the most well established solutions and are used extensively in the geometric optics regime \cite{Jackson1998}. However, pulse solutions to the homogenous Maxwell's equations i.e. representing localised propagation of finite electromagnetic energy, are significantly less well analyzed. A first attempt to produce a mathematical formulation for three dimensional, non-dispersive, source-free solutions to homogenous Maxwell's equations yielded the \textit{focused wave mode} (FWM) solutions, suggesting the possibility for efficient and localized transport of electromagnetic energy in free-space \cite{Brittingham1983}. FWMs were required to meet six criteria: $1)$ satisfy the homogenous Maxwell’s equations, $2)$ be continuous and non-singular, $3)$ have a three-dimensional pulse structure, $4)$ be non-dispersive for all time, $5)$ move at light velocity $c$ along straight lines, and $6)$ carry finite electromagnetic energy. Although the original FWMs were subsequently shown not to violate the sixth criterion \cite{Wu1984,Belanger1984,Sezginer1985}, the finite energy requirement was satisfied by utilizing superpositions of the FWMs over carefully chosen weighting functions \cite{Ziolkowski1985,Ziolkowski1989}. These superpositions are termed as \textit{electromagnetic directed energy pulse trains} (EDEPT’s) and can be tailored so as to give localized propagation of electromagnetic energy in space and time. A wide variety of pulses have been established within the EDEPT family, including the modified power spectrum pulse \cite{Ziolkowski1989}, pulses with azimuthal dependence \cite{Lekner2004,Lekner2004a}, ``focused pancake'' pulses \cite{Feng1998,Feng1999,Hunsche1999}, and the ``\textit{focused doughnut}'' (FD) pulse \cite{Hellwarth1996}. The FD pulse is of particular interest owing to its complex toroidal field geometry, space-time non-separability, and polynomial localisation of energy. Finally, as a free-space toroidal electromagnetic perturbation, investigation of the FD pulse complements the burgeoning field of toroidal electrodynamics in matter.

In this letter we give a detailed description of the propagation properties of FD pulses and their interactions with matter. We demonstrate that due to the toroidal field configuration of the FD pulses, even reflection from dielectric and metallic interfaces can lead to complex field transformations. We show that dielectric nanoparticles under illumination with FD pulses exhibit broadband, multi-mode excitations, including the recently established toroidal response. The paper is organized as follows. Section 2 introduces the theoretical formalism for the description of FD pulses. Studies in the transient domain are considered and these numerical models are used as a basis for examining the interaction of FD pulses with matter. Section \ref{sec_interface} examines the transformation of the FD field topologies when the pulse is incident on dielectric and metallic boundaries, and Section \ref{sec_np} considers the interaction of FD pulses with non-dispersive, dielectric nanoparticles.


\section{The `focused doughnut' pulse }
\label{sec_fd}

\begin{figure}[htbp]
\centering\includegraphics[]{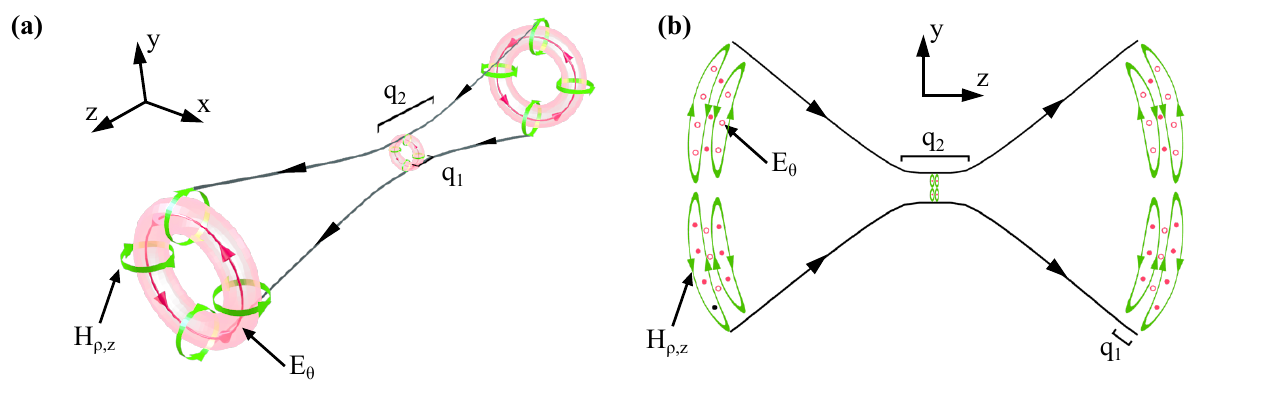}
\caption{Field topology and focusing properties of the "focused doughnut" pulse. The electric $E$ and magnetic $H$ fields are represented by green and red arrows respectively. The effective wavelength parameter $q_1$ and the pulse focal region depth $q_2$ are indicated on both diagrams, along with the region of maximum energy concentration at $z=0$, $t=0$. The pulse envelope is shown by the black lines and arrows.}
\label{FD_artists}
\end{figure}

The FD pulse was first established as a solution to the homogenous Maxwell's equations by Hellwarth and Nouchi \cite{Hellwarth1996}. As space-time non-separable solutions to Maxwell’s equations, FD pulses can be classified in TE and TM field configurations, with the electric and magnetic fields for the TE case given in cylindrical coordinates $(\rho,\theta,z)$ as:
\vspace{0.5cm}

\begin{equation}
E_\theta=-4if_0\sqrt{\frac{\mu_0}{\epsilon_0}}\frac{\rho\left ( q_2-q_1-2ict \right )}{\left [ \rho^2+\left ( q_1+i\tau \right )\left ( q_2-i\sigma \right ) \right ]^3}
\label{Etheta}
\end{equation}

\begin{equation}
H_\rho=4if_0\frac{\rho\left ( q_2-q_1-2iz \right )}{\left [ \rho^2+\left ( q_1+i\tau \right )\left ( q_2-i\sigma \right ) \right ]^3}
\label{Hrho}
\end{equation}

\begin{equation}
H_z=-4f_0\frac{\rho^2-\left ( q_1+i\tau \right )\left ( q_2-i\sigma \right ) }{\left [ \rho^2+\left ( q_1+i\tau \right )\left ( q_2-i\sigma \right ) \right ]^3}
\label{Hz}
\end{equation}
\vspace{0.5cm}

Where $\sigma=z+ct$, $\tau=z-ct$ and $f_0$ is an arbitrary normalisation constant. The parameters $q_1$ and $q_2$ have the dimensions of length and represent respectively the effective wavelength of the pulse and the focal region depth. Beyond the focal region ($|z|>q_2$), the FD diffracts in the same manner as a Gaussian pulse with wavelength $q_1$ and Rayleigh length $q_2$.  The name of this pulse is derived from its three-dimensional field topology, which is illustrated in figure \ref{FD_artists}(a) and (b). The azimuthal electric field in equation \ref{Etheta} forms closed loops that are zero valued on axis. The magnetic field components (equations \ref{Hrho} and \ref{Hz}) form closed loops around the electric field, forming the meridians of a torus structure. The field along the meridians of the torus results in strong longitudinal field component on axis due to the increase in field density within the centre of the torus. The TM solutions are readily obtained by interchanging electric and magnetic field components. Further separating the real and imaginary parts of equations (\ref{Etheta})-(\ref{Hz}) yields two families of pulses, corresponding to a single cycle and $1\frac{1}{2}$ cycle pulse respectively.

\begin{figure}[htbp]
\centering\includegraphics[]{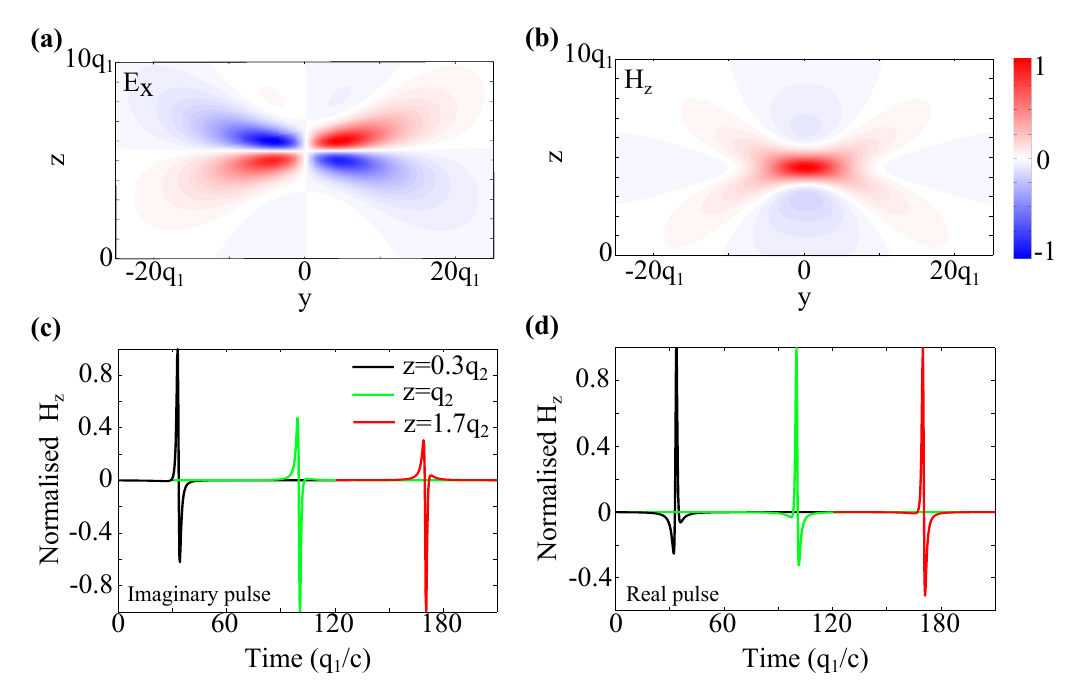}
\caption{Spatial and temporal structure of the "focused doughnut" pulses. Panels (a) and (b) show respectively the normalised transverse electric and longitudinal magnetic fields along a $yz$ cross-section at time $t=45q_1/c$. The characteristic parameters of the FD pulse in all cases are $q_2=100q_1$. All plots are generated from the analytical form of the FD pulse, Eqs. (\ref{Etheta}-\ref{Hz}). The propagation properties of the transverse magnetic pulse are identical to the TE case presented here, by replacing electric (magnetic) with magnetic (electric) fields. (c) and (d) show the evolution of the on-axis field component of both the real and imaginary pulses respectively as they propagate, demonstrating the transformations between the single and $1\frac{1}{2}$ cycle pulses. }
\label{FD_analytic}
\end{figure}

Equations (\ref{Etheta})-(\ref{Hz}) are plotted out explicitly in figure \ref{FD_analytic}(a) and (b) along a $yz$ cross-section, showing explicitly the transverse electric field and longitudinal magnetic field at a time $t=45q_1/c$. The few cycle nature of the pulse evident, with (a) showing a single cycle of the transverse electric field and (b) showing the corresponding $1\frac{1}{2}$ cycle longitudinal magnetic field. As expected, the longitudinal field component of ‘focused doughnut’ pulses is the only component that is non-zero at $\rho=0$. The energy density ($\mu_0H^2+\epsilon_0E^2$) drops off polynomially with $r$ ($r=\sqrt{\rho^2+z^2} $), decaying as $r^{-8}$ (for the real pulse) and $r^{-10}$ (for the imaginary pulse) at the point of maximum focus ($z=0$, $t=0$). All FD solutions have been shown to possess equal and finite total energy \cite{Hellwarth1996}.

It is worth noting that whilst the original paper by Hellwarth and Nouchi categorised the two families of FD into single (imaginary) and $1\frac{1}{2}$ cycle (real) pulses, significant temporal reshaping of the pulses occurs as they propagate along $z$. This is emphasised in figures \ref{FD_analytic}(c) and (d) which shows the spatio-temporal transformations that both the real and imaginary pulses undergo as they propagate along $z$. It is immediately clear from (c) that as the pulses propagate beyond the $q_2$ parameter, the imaginary pulse evolves from single cycle to $1\frac{1}{2}$ cycles, whilst the real pulse evolves from $1\frac{1}{2}$ cycle to single cycle. Similar transformations have been described for other pulses in the EDEPT family and experimentally for single-cycle Gaussian terahertz pulses, and they have been explained in terms of the Gouy phase shift of the pulses \cite{Feng1998,Hunsche1999}.

As a result of their short cycle nature, FD pulses are considered to be ultra--broad bandwidth pulses. Hellwarth and Nouchi give a far-field ($z \gg q_2$) approximation for the Fourier spectra of a real FD in their original paper:
\vspace{0.5cm}

\begin{equation}
F'_\omega=\left ( \frac{i\pi\mu_0f_0\omega\left | \omega \right |sin\varphi }{2rc^2} \right )e^{\frac{i\omega r-\left | \omega \right |Q}{c}}
\label{FDfourier}
\end{equation}

\vspace{0.5cm}
Where $Q=[q_1+q_2-(q_2-q_1)cos\varphi]/2$ and $\varphi$ is the polar angle. The equivalent Fourier spectrum for the imaginary pulse is formed by $F''_\omega=(i\omega/|\omega|)F'_\omega$.

\begin{figure}[t]
\centering\includegraphics[]{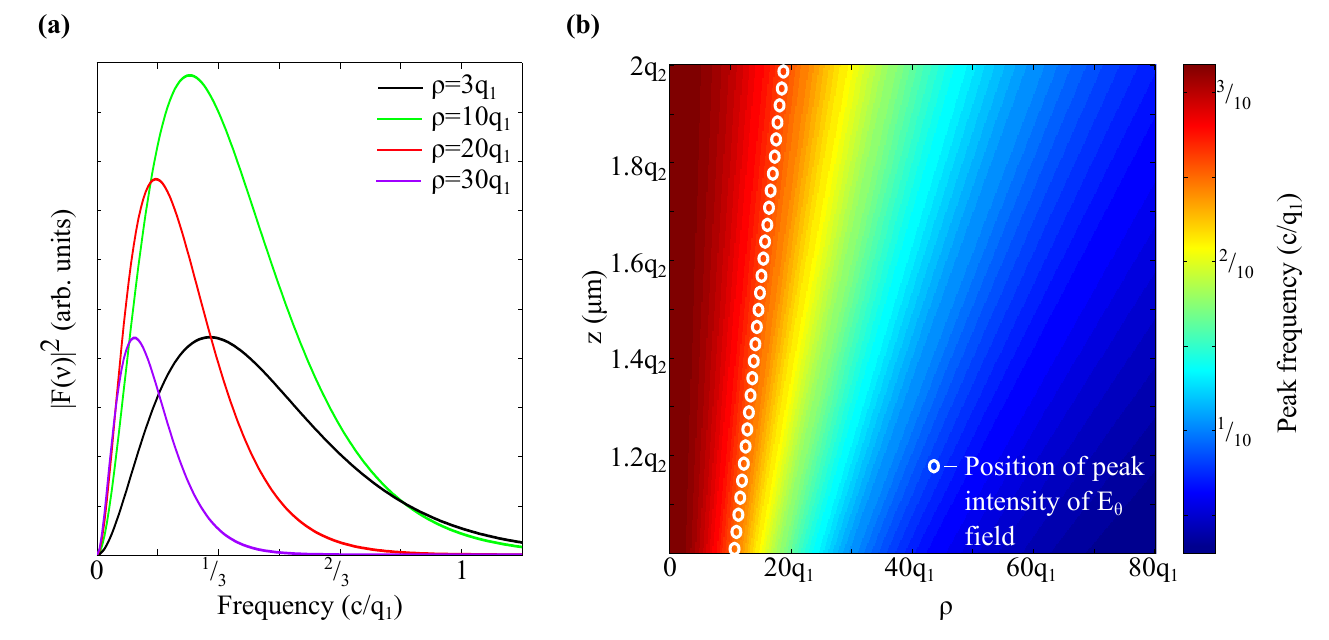}
\caption{Fourier spectrum of the "focused doughnut" pulses. (a) shows the intensity of the Fourier spectrum for $q_2=100q_1$ at $z=q_2$ and for four different radial positions $\rho$. (b) shows the evolution of the peak frequency as a function of $\rho$ and $z$ for $z=q_2 \rightarrow 2q_2$.}
\label{FD_fourier}
\end{figure}

A number of intriguing properties of the FD pulse can be inferred from the Fourier decomposition. The dependence of the Fourier spectrum on $\rho$ is emphasised in figure \ref{FD_fourier}(a) which plots the intensity of the Fourier spectrum at four different values of $\rho$. It can be seen that high frequency components dominate the centre of the FD and lower frequency components become prevalent as $\rho$ increases. This effect can be visualised by considering figure \ref{FD_analytic}(a), where the curvature of the pulse wavefronts indicates a change in frequency as $\rho$ increases. The evolution of the peak frequency as a function of both $\rho$ and $z$ is shown in figure \ref{FD_fourier}(b) over a propagation distance of $q_2$ from $z= q_2 \rightarrow 2q_2$. It can be seen that, as the FD evolves in space, different radial points acquire different peak frequencies. However, it can be noted from the white circles indicating the radial positions of peak intensity, that the peak frequency at the point of peak intensity remains constant as the pulse propagates, in this case at $\sim \frac{c}{4q_1}$ .

The property of a varying peak frequency transverse to the pulse propagation direction is known as spatial chirp, and is a common occurrence in ultrafast optics. However, whilst the spatial chirp of the FD is intrinsic to the pulse, it generally arises in ultrafast optics due to misalignment of optical elements used for production of ultrashort pulses e.g. prisms, tilted substrates and Fourier pulse shapers \cite{Gu2004}. This lack of control of the phenomenon leads to the spatial chirp being considered an undesirable side-effect. The well-defined spatial chirp of the FD pulse however, allows to exploit this property by coupling frequency information to spatial positions of the pulse, a situation which is of interest for spectroscopy for example. In addition, it can be noted that the spatial chirp of the FD is axially symmetric, as per the topology of the pulse, and so can in fact be considered as radial chirp.

A further point is that for all $\rho$, the bandwidth of the FD pulse is greater than the peak frequency. Consider for instance the Fourier spectrum at $\rho=10q_1$ (green curve in figure \ref{FD_fourier}(a)). The peak frequency $\nu_0$ at this $\rho$ value is $\frac{c}{4q_1}$, compared to a full width at half maximum bandwidth of $\sim 1.1\nu_0$. This can be taken in contrast to the typical bandwidth-limited pulses produced by solid-state laser, for which the bandwidth will always be smaller than the peak frequency.

The nature of the FD pulses has resulted in several potential applications being presented. The strongly localised nature of the pulse for $|z|<q_2$ suggests microscopy, communications and directed energy transfer as potential uses \cite{Hellwarth1996}. The presence of the longitudinal field component was discussed in detail by Hellwarth and Nouchi in terms of a mechanism for accelerating co-propagating particles, and later by Varin et al. \cite{Varin2005}.


\section{Interaction with plane interfaces}
\label{sec_interface}

\begin{figure}[htbp]
\centering\includegraphics[]{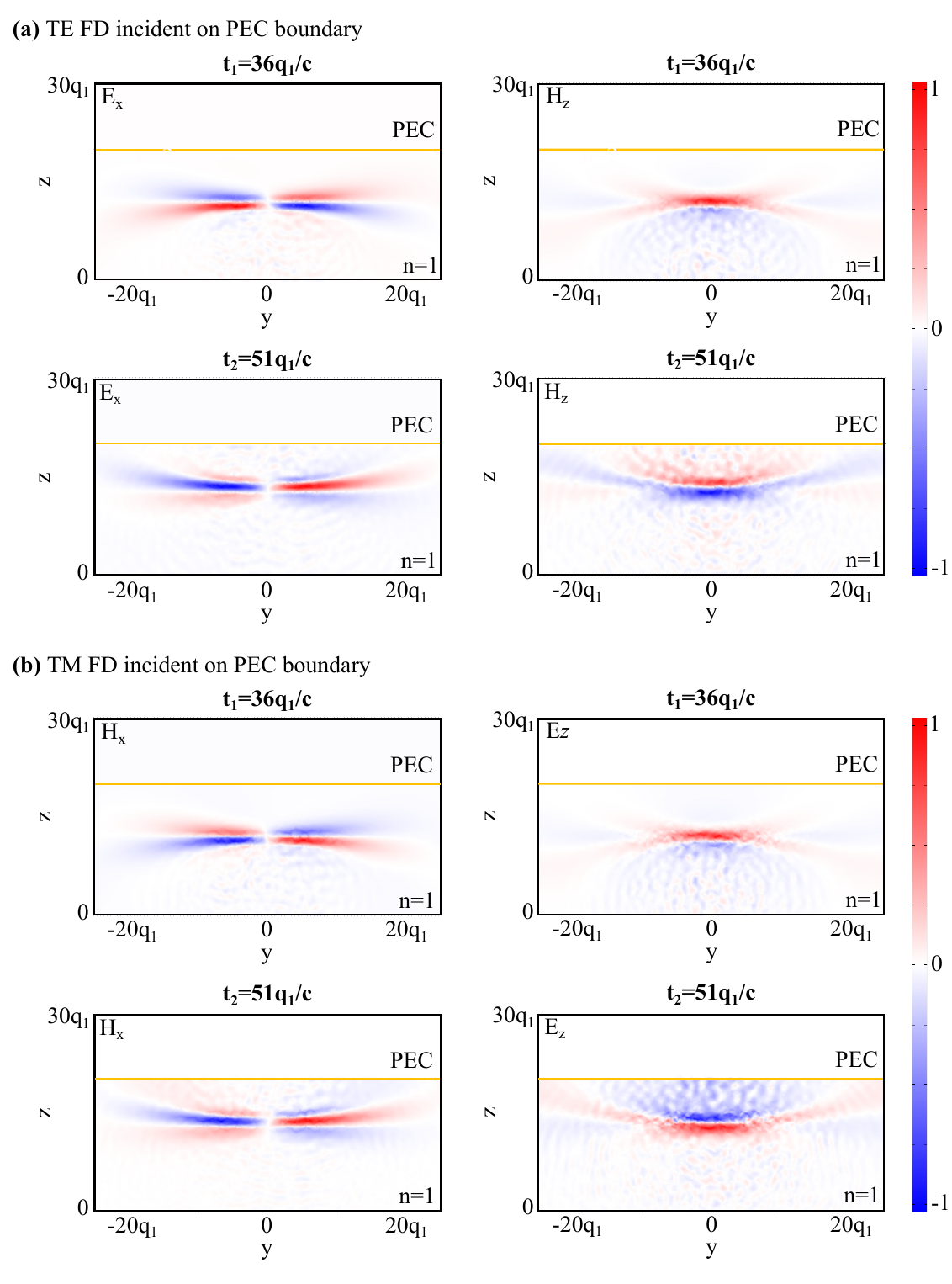}
\caption{Reflection of FD pulses from a perfect conductor. (a) Transverse electric (left) and longitudinal (right) magnetic field components of a transverse electric FD pulse before ($t_1$) and after ($t_2$) reflection. (b) Similar to (a) but for a transverse magnetic pulse. In both cases the parameters of the FD pulse are $q_2=100q_1$ and the boundary is located at a distance $z=20q_1$ from the focal point of the pulse ($z=0$). All field components have been normalized to their maximum value.}
\label{FD_interface_PEC}
\end{figure}

\begin{figure}[htbp]
\centering\includegraphics[]{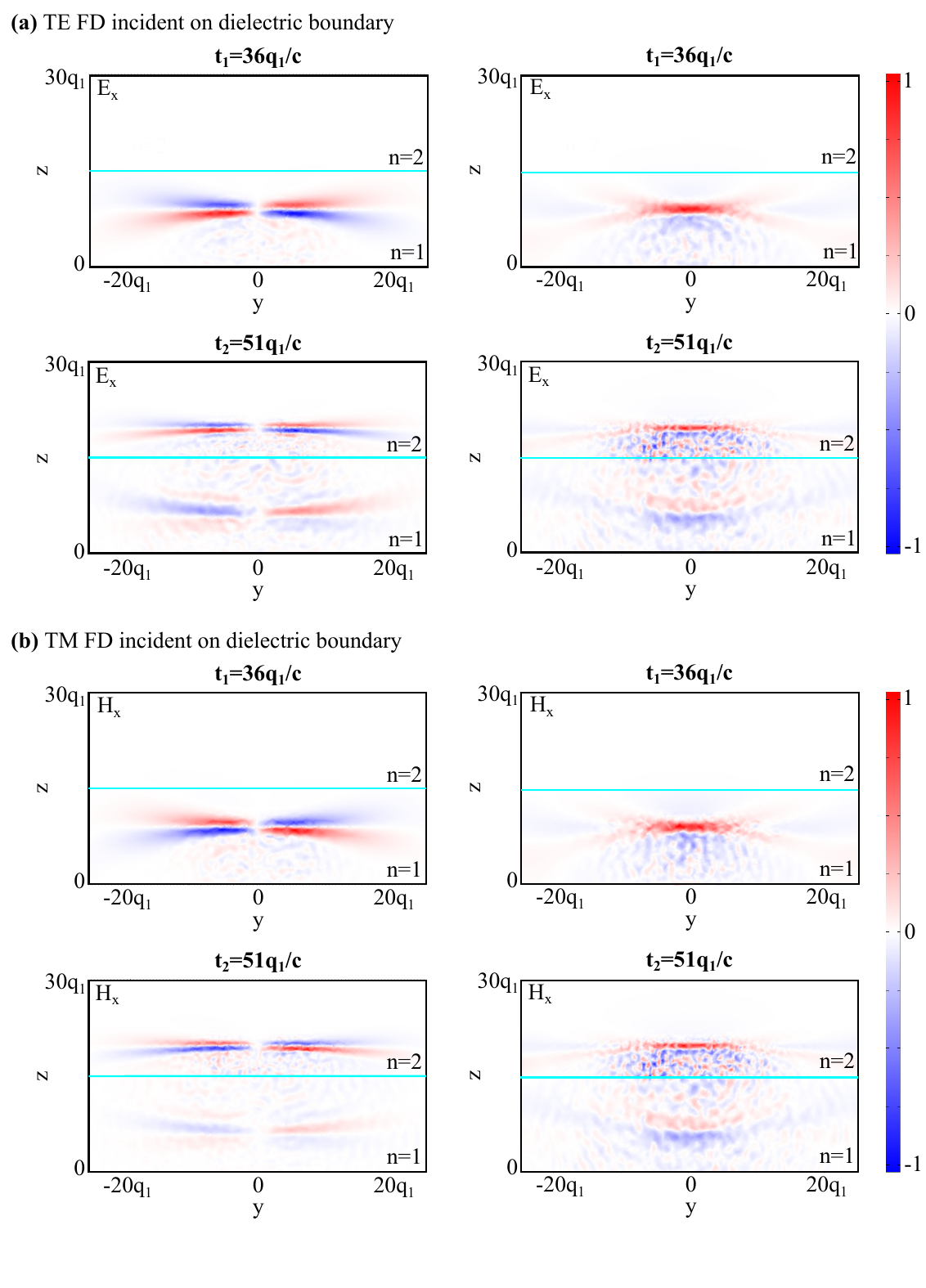}
\caption{Reflection and refraction of "focused doughnut" pulses at a vacuum-dielectric interface. (a) Transverse electric (left) and longitudinal magnetic field (right) components of a transverse electric "focused doughnut" pulse before ($t_1$) and after ($t_2$) incidence on the interface. (b) Similar to (a) but for a transverse magnetic pulse. In both cases the parameters of the "focused doughnut" pulse are $q_2=100q_1$ and the boundary is located at a distance $z=15q_1$ from the focal point of the pulse ($z=0$). All field components have been normalized to their maximum value. The dielectric is considered to be semi-infinite with a refractive index $n=2$.}
\label{FD_interface_dielec}
\end{figure}

Although the properties of the FD pulse in freespace have been well established, and the propagation dynamics and reshaping of similar pulses have been described in the literature, there has been a very limited treatment of the interaction of such complex pulses with matter. In general, the problem is not amenable to analytical considerations. To this end, we employ a commerical finite element solver package to evaluate the interactions of the FD pulse with continuous matter. In this section, we give the first explicit analysis of the transformations of the FD field components under reflection and transmission when incident on semi-infinite dielectrics and \textit{perfect electric conductor} (PEC) boundaries.

We first consider a PEC boundary located at $z=20q_1$ in the free-space propagation domain. Figure \ref{FD_interface_PEC}(a) and (b) show the TE and TM pulse respectively at two times -- one prior to incidence on the boundary ($t_1$) and one after the pulse has been reflected ($t_2$). For the TM FD pulse in figure \ref{FD_interface_PEC}(b), the transverse magnetic and longitudinal electric field components are shown. It is clear from examining the field distributions at the two time steps, that reflection at the boundary results in the reversal of both the electric and magnetic field components. After reflection, the transverse magnetic fields are counter-rotating with respect to the propagation direction. The longitudinal electric field component is dominated by a strong component parallel to the propagation direction at the pulse front. Upon reflection, the electric field at the pulse front is anti-parallel to the propagation of the pulse. The presence of a longitudinal field component, anti-parallel to the propagation direction, at the leading edge of the pulse is a particularly intriguing and non-intuitive property of the reflected TM FD pulse.

In contrast, the TE FD pulse incident on the PEC boundary in figure \ref{FD_interface_PEC} (a) does not undergo any field reversals upon reflection. After reflection, the transverse electric field of the pulse rotates in the same direction with respect to the propagation direction as before reflection. Similarly, the longitudinal magnetic field component both before and after reflection has a component parallel to the propagation direction leading the pulse.

This modelling of the FD pulse also highlights the spatio-temporal transformations the pulse undergoes, as described in Section \ref{sec_fd} and figure \ref{FD_analytic} (c) and (d). It can be clearly seen in figure \ref{FD_interface_PEC} (a) how, after reflection at the boundary, the transverse magnetic field of the pulse is beginning to transform from single cycle to $1\frac{1}{2}$ cycle nature. Equivalently, the longitudinal electric field is beginning to transform from $1\frac{1}{2}$ cycle to single cycle nature. This transformation is also evident in figure \ref{FD_interface_PEC} (b) for the TE pulse.

We now consider interaction at a dielectric boundary. The modelling space is separated into two regions -- one free space and one of refractive index $n=2$, with the boundary located at $z=15q_1$ so as to illustrate both the reflected and transmitted pulse. Figure \ref{FD_interface_dielec} shows the results of these models in both TE and TM incidence cases, with the transverse and longitudinal fields shown at two times -- one prior to incidence on the boundary ($t_1$) and one after the pulse has been reflected and transmitted ($t_2$). In both polarisation cases the toroidal topology of the pulse is maintained after being transmitted though the dielectric boundary and it undergoes the expected increase in momentum within the medium as for conventional electromagnetic pulses. Similarly the reflected pulse also maintains its toroidal topology. Evaluation of transmission and reflection coefficients for the reflected and transmitted pulses indicates that both TE and TM FD pulses interact with the semi-infinite dielectric as predicted by the Fresnel equations.

It is worth noting that all models in this section utilise idealised non-dispersive metals and dielectrics, in the form of PEC and a dielectric refractive index of $n=2$. This is due to limitations in Maxwell's equations solver utilised for this study, which prohibits temporal dispersion for transient models. As illustrated previously in this paper, the FD pulse is highly broadband with a bandwidth greater than the peak frequency. For realistic materials, it is likely that dispersive effects would be present over such a wide frequency range, inducing some reshaping to the temporal profile of the pulse. However, it is expected that this will not limit the analysis of the reflected and transmitted geometries. It could also be considered that this would be a valid description for an FD pulse in the microwave regime, in which metals can generally be approximated as PEC for thicknesses greater than $\mu$m-scale \cite{Youngs2005,Youngs2006}.


\section{Interaction with nanoparticles}
\label{sec_np}

\begin{figure}[htbp]
\centering\includegraphics[]{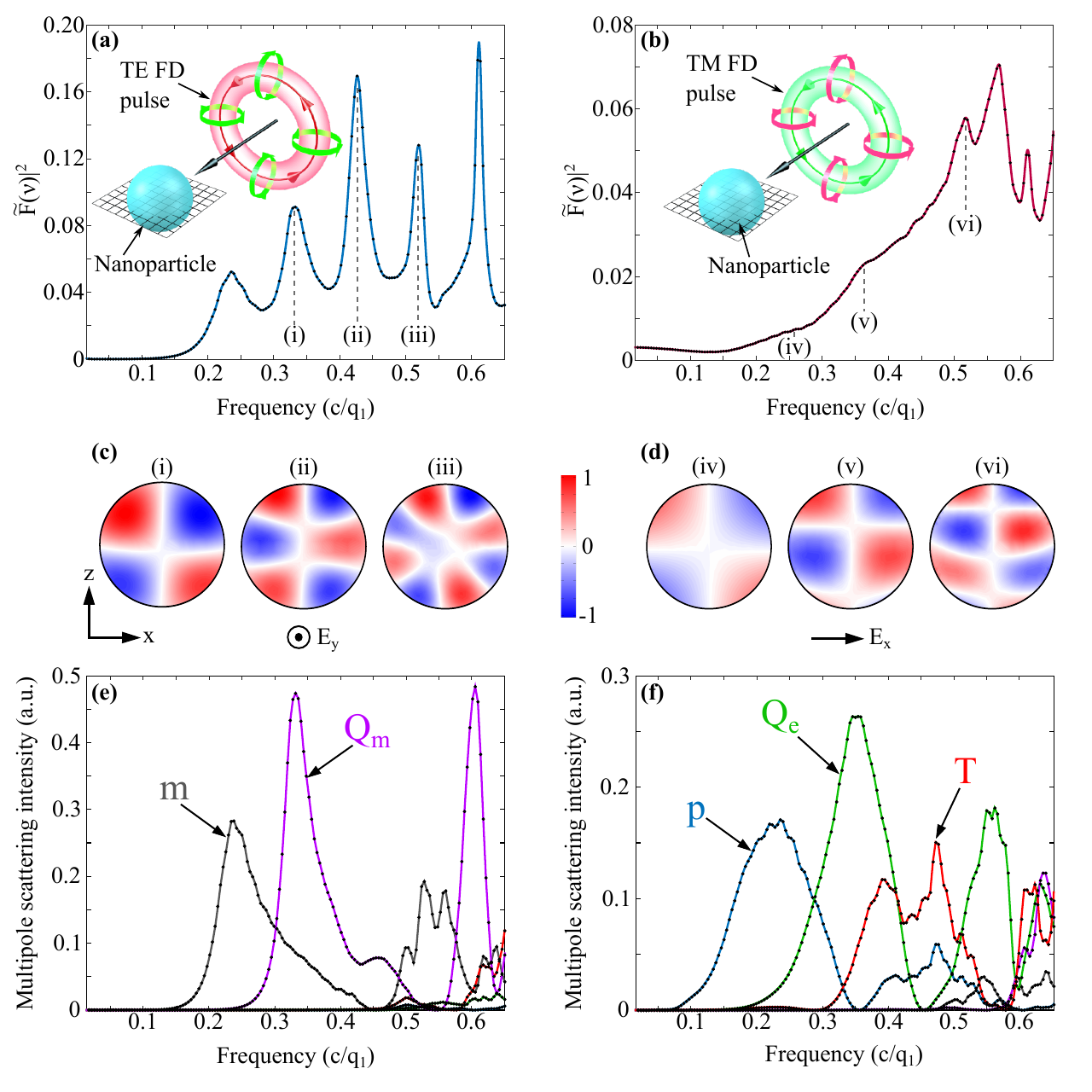}
\caption{Interactions of "focused doughnut" pulses with spherical dielectric nanoparticles. (a) \& (b) show the electric field intensity integrated over the volume of the nanoparticle as a function of frequency under excitation with a transverse electric (TE) and a transverse magnetic (TM) "focused doughnut" pulse, respectively. (c) \& (d) show the electric field distributions on an $xz$ cross-section of the nanoparticle (see grid in the insets to (a) \& (b)) at resonance positions (i)-(vi). (e) \& (f) show the cartesian multipole expansion up to quadrupole order (electric dipole $p$, magnetic dipole $m$, toroidal dipole $T$, electric quadrupole $Q_e$, and magnetic quadrupole $Q_m$) for illumination with TE and TM "focused doughnut" pulses, respectively. In (a-b) \& (e-f) dots correspond to simulation data points, while lines serve as eye guides.    }
\label{FD_NP}
\end{figure}

The space-time non-separable and ultra-broadband nature of the FD pulses is expected to manifest in an interesting manner when considering their interaction with dielectric nanoparticles. The case considered is that of a spherical nanoparticle located at $\rho=z=0$. The radius of the nanoparticle is given as $q_1$, such that it is less than the width of the FD pulse. In this regime, excitation by the ultra-broadband FD pulse can be expected to induce multiple Mie modes of the dielectric nanoparticle. As in the previous section, the dielectric is given a non-dispersive refractive index of $n=2$. For an incident TM FD pulse, the interaction with the nanoparticle is dominated by the longitudinal electric field on axis. A schematic of the interaction is shown in the insets of figure \ref{FD_NP}(a) and (b), for the TE and TM case respectively.

We first evaluate the normalised electric field intensity within the nanoparticle as a function of frequency. These are shown in figure \ref{FD_NP}(a) and (b) for the TE and TM incidence case respectively. For TE FD incidence, a clear series of Mie modes are excited corresponding to resonant distributions of the azimuthal field throughout the nanoparticle. This is emphasised in figure \ref{FD_NP}(c), which shows out-of-plane electric field distributions for a cross-section through the nanoparticle at three different peaks.  In contrast, the spectrum for excitation of the nanoparticle by a TM pulse is more complicated, owing to the interplay between the radial and longitudinal electric field components. As such, the TM excitations are considerably weaker than those from TE pulse incidence. Modes corresponding to distributions of $x$-aligned (in-plane) $E$ field from three frequencies are shown in figure \ref{FD_NP}(d), corresponding to a series of Mie resonances, similar to the TE pulse case.

To extend this description of FD-nanoparticle interaction, we evaluate the scattering from the microscopic mutlipoles excited within the nanoparticle. Owing to the toroidal topology of the FD pulse, the toroidal family of multipoles are included in this analysis. In recent years, the previously elusive toroidal dipole has been demonstrated as a dominant contributor to the scattering of mutliple systems with toroidal topology \cite{Kaelberer2010,Fedotov2013,Basharin2015}. This includes systems composed entirely of dielectric elements \cite{Basharin2015}. In contrast, here the incident excitation (the FD) is in possession of toroidal topology, whereas the excited system of the nanoparticle is not. This is particularly relevant for the TM FD pulse, where its azimuthal $H$ and radial and longitudinal $E$ fields are analagous to the poloidal currents and closed loop of magnetic field that compose a toroidal dipole. As such, it is anticipated that a TM FD should excite a non-negligible toroidal dipole within the nanoparticle.

In the case of TE FD incidence on the nanoparticle (figure \ref{FD_NP}(e)) it is clear that the dominant contributors to the scattering are the magnetic multipoles. This is anticipated owing to the azimuthal $E$ field configuration of the TE FD pulse. Contributions from the electric and toroidal multipoles are significantly suppressed. For the case of TM incidence  (figure \ref{FD_NP}(f)) however, the mutlipole excitations are more complex. As expected, electric multipoles dominate at lower frequencies, as a result of coupling to the longitudinal $E$ field of the incident pulse. However at $\sim0.46(c/q_1)$, the toroidal dipole becomes the dominant scattering multipole up to quadrupole order. This is a particularly intriguing feature as it demonstrates a significant toroidal response in a system with non-toroidal geometry, reiterating the importance of toroidal multipoles in electrodynamics. It is anticipated that different topologies of nanoparticle interacting with FD pulses could show an even more significant presence of toroidal multipoles.

\section{Summary}

In conclusion, we have demonstrated that the field configuration of FD pulses, consisting of longitudinal and circulating azimuthal fields, undergoes complex polarisation-sensitive transformations under reflection and refraction at metallic and dielectric interfaces. When FD pulses interact with (non-dispersive) dielectric particles, the single-cycle, broadband nature of the pulse results in the resonant excitation of multiple Mie modes over a wide frequency region. Finally, the toroidal topology of the pulse allows to excite a non-negligible toroidal dipole in a system of non-toroidal topology. Our results highlight the potential of the FD pulses, especially within the context of the nascent field of toroidal electrodynamics.

\section*{Acknowledgements}

The authors would like to acknowledge the financial support of the Defence Science and Technology Laboratory, Engineering and Physical Sciences Research Council (U.K.), the Leverhulme Trust and the Royal Society.



\end{document}